\documentclass[useAMS,usenatbib]{mn2e}
\usepackage{graphicx}

\title[Gamma rays from molecular clouds illuminated by cosmic rays]{Gamma rays from molecular clouds illuminated by cosmic rays escaping from interacting supernova remnants}
\author[Y. Ohira et al.]{Yutaka Ohira$^{1}$\thanks{E-mail:ohira@post.kek.jp}, Kohta Murase$^{2},^{3}$ and Ryo Yamazaki$^{4}$\\
$^{1}$Theory Center, Institute of Particle and Nuclear Studies, KEK, 1-1 Oho, Tsukuba 305-0801, Japan\\
$^{2}$Department of Physics, Tokyo Institute of Technology, 2-12-1, Ookayama, Meguro-ku, Tokyo 152-8550\\
$^{3}$Center for Cosmology and AstroParticle Physics, 191, W. Woodruff. Ave., Columbus, Ohio 43210, USA\\
$^{4}$Department of Physics and Mathematics, Aoyama Gakuin University, 
5-10-1 Fuchinobe, Sagamihara 252-5258, Japan}
\begin{document}


\pagerange{\pageref{firstpage}--\pageref{lastpage}} \pubyear{2010}

\maketitle

\label{firstpage}

\begin{abstract}
Recently, gamma-ray telescopes {\it AGILE} and {\it Fermi} observed several middle-aged supernova remnants (SNRs) interacting with molecular clouds.
It is likely that  their gamma rays arise from the decay of neutral pions produced by the inelastic collision between cosmic rays (CRs) and nucleons, which suggests that SNRs make the bulk of Galactic CRs.
In this paper, we provide the analytical solution of the distribution of CRs that have escaped from a finite-size region, which naturally explains observed broken power-law spectra of the middle-aged SNRs.
In addition, the typical value of the break energy of the gamma-ray spectrum, 1--10~GeV, is naturally explained from the fact that the stellar wind dynamics shows the separation between the molecular clouds and the explosion center of about 10~pc.
We find that a runaway-CR spectrum of the four middle-aged SNRs (W51C, W28, W44 and IC~443) interacting with molecular clouds could be the same, even though it leads to different gamma-ray spectra. 
This result is consistent with that of recent studies of the Galactic CR propagation, and supports that SNRs are indeed the sources of Galactic CRs.

\end{abstract}


\section{Introduction}

Supernova remnants (SNRs) are thought as the origin of Galactic cosmic rays (CRs). 
The most popular acceleration mechanism at SNR is the diffusive shock acceleration (DSA) \citep{axford77,krymsky77,bell78,blandford78}.
To make the Galactic CRs, a few tens of percent of the explosion energy of a supernova is converted to the CR energy \citep{baade34}.
Recent gamma-ray telescopes, such as {\it Fermi} and {\it AGILE}, 
suggest that middle-aged SNRs interacting with molecular clouds likely emit hadronic gamma-rays,
which arise from the decay of $\pi^0$ produced in inelastic collisions of the accelerated protons with molecular clouds
\citep{abdo09b,abdo10a,abdo10b,abdo10c,tavani10,giuliani10}.
This fact supports that the SNRs make the bulk of the Galactic CRs.
{\it Fermi} observations revealed that the gamma-ray spectra of the middle-aged SNRs
have a broken power law form,  that the break photon energies are typically $\sim$1--10~GeV, 
and that the spectral indices above the break energy are different from each other.

Several theoretical interpretations of the gamma-ray observations of middle-aged SNRs have been proposed.
\citet{aharonian96} and \citet{gabici09} have investigated the gamma-ray spectrum from molecular clouds illuminated by CRs which have escaped from a nearby SNR. 
They showed that the spectrum is steeper than predicted by DSA at the shock because of the energy dependent diffusion.
In addition, \citet{aharonian96} showed that the spectrum has a broken power law form due to the finite size of the emission region.
On the other hand, \citet{malkov10} recently proposed a possible explanation. 
They claimed that in the partially ionized medium, CRs accelerated at the shock has a broken power-law spectrum --- above the break energy, the damping of waves that resonantly scatter the CRs is significant, making the steeper spectrum than below the break energy.

In this paper, we propose another explanation for the observed middle-aged SNRs interacting with molecular clouds, by reinvestigating the distribution of CRs that have escaped from an SNR.
So far, the point source approximation has been widely adopted.
However, for SNRs interacting with molecular clouds, the extension of the CR sources is important.
In section~2, taking into account the effect of a finite-size source, we first provide the distribution of CRs that have escaped from a SNR.
Then, we consider the dynamical evolution of CRs interacting with molecular clouds, because the CR spectrum that has escaped from an SNR depends on the time evolution of the maximum energy \citep{ohira10}.
In section~3 and section~4, we provide two models of the evolution of the maximum energy.
The gamma-ray spectra from molecular clouds are calculated in section~5.
Section~6 is devoted to a discussion.

\section[]{Distribution of CRs escaping from a SNR}
\label{sec:2}
In this section, 
we derive the distribution of the runaway CRs at a given distance $r$ from the SNR center 
and at a given time $t$ from the explosion time, $f(t,r,p)$, where $p$ is the CR momentum.
We here assume that the system is spherically symmetric and the diffusion coefficient in the interstellar 
medium $D_{\rm ISM}(p)$ is spatially uniform and depends on the CR momentum.
Then, we solve the diffusion equation given by
\begin{equation}
\frac{\partial f}{\partial t}(t,r,p) - D_{\rm ISM}(p)\Delta f(t,r,p) = q_{\rm s}(t,r,p)~~,
\label{eq:diffusion}
\end{equation}
where $q_{\rm s}$ is the source term of CRs.
Considering the escape process \citep{ptuskin05,ohira10}, the CRs with a momentum $p$ escape from an 
SNR at $t=t_{\rm esc}(p)$.  
The Green function, that is, the solution in 
the point source case, {\bf $q_{\rm s}=N_{\rm esc}(p)\delta({\bf r})\delta(t-t_{\rm esc}(p))$}, is \citep{atoyan95}
\begin{equation}
f_{\rm point}(t,r,p)=\frac{e^{-\left(\frac{r}{R_{\rm d}(t,p)}\right)^2}}{\pi^{3/2}R_{\rm d}(t,p)^3}N_{\rm esc}(p)~~,
\label{eq:fpoint}
\end{equation}
where 
\begin{equation}
R_{\rm d} = \sqrt{4D_{\rm ISM}(p)(t-t_{\rm esc}(p))}~~,
\label{eq:rd}
\end{equation}
and 
\begin{equation}
N_{\rm esc}(p) = \int {\rm d}t\int{\rm d}^3r~q_{\rm s}(t,r,p)~~.
\end{equation}
Now, taking into account the fact that the CRs escape from the SNR surface \citep{ohira10}, the source term is replaced with
\begin{equation}
q_{\rm s}=\frac{N_{\rm esc}(p)}{4\pi r^2} \delta(r-R_{\rm esc}(p)) \delta(t-t_{\rm esc}(p))~~,
\end{equation}
where $R_{\rm esc}(p)$ is the radius where the CRs with momentum $p$ escape from the SNR.
Then, we find the solution to equation~(\ref{eq:diffusion}) as
\begin{eqnarray}
f_{\rm ext}(t,r,p)&=&\int {\rm d}^3r'  f_{\rm point}(t,|{\bf r}-{\bf r'}|,p) \frac{\delta(r'-R_{\rm esc}(p))}{4\pi r'^2} \nonumber \\
&=&\frac{e^{-\left(\frac{r-R_{\rm esc}(p)}{R_{\rm d}(t,p)}\right)^2}-e^{-\left(\frac{r+R_{\rm esc}(p)}{R_{\rm d}(t,p)}\right)^2}}{4\pi^{3/2}R_{\rm d}(t,p)R_{\rm esc}(p)r}N_{\rm esc}(p)~~.
\label{eq:fext}
\end{eqnarray}
Note that $f_{\rm ext}$ has a spectral break at $p=p_{\rm br,ext}$ because $f_{\rm ext}\approx N_{\rm esc}/4\pi^{3/2} R_{\rm d}R_{\rm esc}r$ for $r\sim R_{\rm esc}>R_{\rm d}$,
otherwise $f_{\rm ext} \approx f_{\rm point}$, where $p_{\rm br, ext}$ is obtained from
\begin{equation}
R_{\rm esc}(p_{\rm br,ext}) = R_{\rm d}(p_{\rm br,ext})~~.
\end{equation}
Below $p_{\rm br, ext}$ ($R_{\rm esc}>R_{\rm d}$), the effect of a finite-size source becomes important.

The observed gamma-ray spectrum should be calculated by the volume-integrated CR spectrum.
In the case of the SNR-molecular-cloud interacting system, the emission region is a dense cloud or cavity wall, whose density is high in the region $L_1\leq r\leq L_2$.

Considering the stellar wind before the supernova explosion, the inner radius of the molecular cloud $L_1$ is about a few tens of pc \citep{weaver77}.
The volume-integrated spectrum of CRs in the region $L_1\leq r\leq L_2$ at the SNR age $t_{\rm age}$ is calculated as
\begin{equation}
F(p)=\int_{L_1}^{L_2} f_{\rm ext}(t_{\rm age},r,p)4\pi r^2{\rm d}r ~~.
\end{equation}
Substituting Eq.~(\ref{eq:fext}) into the above, we obtain
\begin{eqnarray}
F(p)&=& \frac{N_{\rm esc}(p)}{2} \left\{ \frac{R_{\rm d}}{\sqrt{\pi}R_{\rm esc}} \left( e^{- \left( \frac{L_1-R_{\rm esc}}{R_{\rm d}} \right)^2} 
-e^{- \left( \frac{L_2-R_{\rm esc}}{R_{\rm d}} \right)^2} \right. \right. \nonumber \\
&&\left.-e^{- \left( \frac{L_1+R_{\rm esc}}{R_{\rm d}} \right)^2}
+e^{- \left( \frac{L_2+R_{\rm esc}}{R_{\rm d}} \right)^2} \right)  \nonumber \\
&&+{\rm erf}\left(\frac{L_2-R_{\rm esc}}{R_{\rm d}}\right)
-{\rm erf}\left(\frac{L_1-R_{\rm esc}}{R_{\rm d}}\right)  \nonumber \\
&&\left.+{\rm erf}\left(\frac{L_2+R_{\rm esc}}{R_{\rm d}}\right)
-{\rm erf}\left(\frac{L_1+R_{\rm esc}}{R_{\rm d}}\right) \right\},
\label{eq:fp}
\end{eqnarray}
where ${\rm erf}(x)=(2/\sqrt{\pi})\int_0^{x}e^{-y^2}{\rm d}y$ is the error function.

There are four characteristic momentum regimes determined by the comparison between the distances that CRs can reach, $R_{\rm esc}(p)+R_{\rm d}(p)$ and $L_{1,2} (\ge R_{\rm esc}(p))$, and by the comparison between $R_{\rm d}$ and $R_{\rm esc}$; 
\begin{enumerate}
  \item $L_{1,2} <R_{\rm esc}(p)+R_{\rm d}(p)~{\rm and}~R_{\rm esc}<R_{\rm d}$,
  \item $L_{1,2}<R_{\rm esc}(p)+R_{\rm d}(p)~{\rm and}~R_{\rm esc}>R_{\rm d}~(L_1 \sim R_{\rm esc})$,
  \item $L_1<R_{\rm esc}(p)+R_{\rm d}(p)<L_2$,
  \item and $R_{\rm esc}(p)+R_{\rm d}(p)<L_{1,2}$.
\end{enumerate}
%
Let us define $p_{\rm br,2}$ and $p_{\rm cut}$ as  the solutions to the following equations: 
\begin{eqnarray}  
&&L_2 = R_{\rm esc}(p_{\rm br,2})+R_{\rm d}(p_{\rm br,2})\label{eq:condition2} \\
&&L_1 = R_{\rm esc}(p_{\rm cut})+R_{\rm d}(p_{\rm cut})\label{eq:condition3}~~.
\end{eqnarray}
We may expect $p_{\rm cut}<p_{\rm br,2}$ as long as the momentum dependence of $R_{\rm esc}$ is weak enough and/or $R_{\rm d}$ is more important. Then, a significant fraction of CRs with energies above $p_{\rm br,2}$ leave the molecular cloud, while CRs below $p_{\rm cut}$ do not essentially reach the cloud.

Furthermore, we approximate $e^{-x^2}\approx1-x^2+x^4/2$ and ${\rm erf}(x)\approx (2/\sqrt{\pi})(x-x^3/3)$ for $0<x<1$, and $e^{-x^2}\approx 0$ and ${\rm erf}(x)\approx 1$ for $1<x$.
Then,  for $p_{\rm br,ext}>p_{\rm br,2}$, we obtain
\begin{eqnarray}
F(p)\propto N_{\rm esc}(p) \times \left\{ \begin{array}{ll}
R_{\rm d}(p)^{-3}& (p>p_{\rm br,ext}) \\
R_{\rm d}(p)^{-1}R_{\rm esc}(p)^{-1} & (p_{\rm br,ext}>p>p_{\rm br,2}) \\
p^0  & (p_{\rm br,2}>p> p_{\rm cut}) \\
0  & (p_{\rm cut}> p)
\end{array} \right. ~~.
\label{eq:f}
\end{eqnarray}
For $p_{\rm br,2}>p_{\rm br,ext}$, we obtain
\begin{eqnarray}
F(p)\propto N_{\rm esc}(p) \times \left\{ \begin{array}{ll}
R_{\rm d}(p)^{-3}& (p>p_{\rm br,2}) \\
p^0  & (p_{\rm br,2}>p> p_{\rm cut}) \\
0  & (p_{\rm cut}> p)
\end{array} \right. ~~.
\label{eq:f1}
\end{eqnarray}
Note that there is no break at $p_{\rm br,ext}$ since the effect of a finite-size source is smeared by spatial integration.

The spectral breaks at $p_{\rm br,ext}$ and $p_{\rm br,2}$ come from the finiteness of  the source and the emission regions, respectively.
To estimate the values of $p_{\rm br,2}$ and $p_{\rm cut}$, we first assume the diffusion coefficient of the interstellar medium
\begin{equation}
D_{\rm ISM}(p) = 10^{28}~\chi \left(\frac{cp}{10{\rm GeV}}\right)^{\delta}~{\rm cm^2~s^{-1}}~~,
\label{eq:dism}
\end{equation}
where $\chi$ is constant. 
CR propagation models require $\chi\sim1$ and $\delta\sim 0.5$ as the galactic average value \citep{berezinskii90}.
Assuming that $R_{\rm esc}$ and $t_{\rm esc}$ are constant with $p$ (because $p$-dependence of $R_{\rm esc}$ is weak), the typical value of $p_{\rm cut}$ is estimated as 
\begin{equation}
p_{\rm cut} = 7~\left(\frac{\chi}{1} \right)^{-1/\delta} \left(\frac{L_1-R_{\rm esc}}{5~{\rm pc}} \right)^{2/\delta} \left(\frac{t-t_{\rm esc}}{10^4~{\rm yr}} \right)^{-1/\delta} ~{\rm GeV}/c ~~,
\end{equation}
where we assume $\delta=0.5$, and $p_{\rm br,2}$ is obtained by replacing $L_1$ with $L_2$.
Although $L_1$ is order of $10~{\rm pc}$, {\it AGILE} observations show that a few$\times 100~{\rm MeV}$ photons come from molecular clouds interacting with the SNR, that is, CRs with a few GeV energies have to reach the molecular clouds.
In order for $p_{\rm cut}$ to be smaller than $\sim$1~GeV/$c$, $R_{\rm esc}$ should be order of $L_1$.
Therefore, the effect of the finite-size source considered here is important.
To estimate the values of those momentum breaks, we need to specify a model of $R_{\rm esc}(p)$. However, properties of particle escape depend on models. As we see in section~4, when the whole SNR shell interacts with the molecular cloud, another momentum break $p_{\rm br,1}$ should be introduced, which plays a more crucial role than $p_{\rm cut}$.

Given $R_{\rm esc}(p)$, $t_{\rm esc}(p)$, and $N_{\rm esc}(p)$, one can obtain $f_{\rm ext}(t,r,p)$ and $F(p)$.
In section~3 and section~4, we provide two models fixing $R_{\rm esc}(p)$, $t_{\rm esc}(p)$,  and $N_{\rm esc}(p)$ in the context of SNR-molecular cloud interacting system.

\section[]{Model 1: SNR interacting with a small molecular cloud}
\label{sec:3}
In this section, we consider the case in which a small part of a SNR shell interacts with a molecular cloud.
In this case, the molecular cloud affect neither dynamical evolution of the system nor the shock environments to confine CRs, so that
both $R_{\rm esc}(p)$ and $N_{\rm esc}(p)$ are those for an isolated SNR which are obtained from \citet{ohira10}. 
Here we briefly summarize the results of \citet{ohira10}.
In the framework of DSA, CRs are scattered by the turbulent magnetic field and go back and forth across the shock front.
Once CRs reach far upstream where the shock front can not be identified as a plane, the CRs can not go back to the shock front, that is, the CRs escape from the SNR. 
In the context of DSA, the diffusion length of CR is $D(p)/u_{\rm sh}$, where $D(p)$ and $u_{\rm sh}$ are the diffusion coefficient in the vicinity of the shock and the shock velocity, respectively.
The momentum of escaping CRs, $p_{\rm esc}$, is determined by
\begin{equation}
\frac{D(p_{\rm esc})}{u_{\rm sh}}\sim \ell_{\rm esc}~~,
\label{eq:pesc0}
\end{equation}
where $\ell_{\rm esc}$ is the distance of the escape boundary from the shock front.
We assume the Bohm like diffusion coefficient $D=D_0 p$
and $\ell_{\rm esc}=\kappa R_{\rm sh}$ as a geometrical confinement condition, 
where we adopt $\kappa=0.04$  throughout the paper \citep{ptuskin05}.
Then, we obtain
\begin{equation}
p_{\rm esc} = \kappa D_0^{-1}R_{\rm sh}u_{\rm sh}~~.
\label{eq:pesc1}
\end{equation}
Although $D_0$ has not been understood in detail, $p_{\rm esc}$ starts to decrease when the Sedov phase begins at $t=t_{\rm Sedov}$.
Hence we assume that CRs with the knee energy, $cp_{\rm knee}=10^{15.5}~{\rm eV}$, escape at $t=t_{\rm Sedov}$.
We adopt a phenomenological approach based on the power-law dependence,
\begin{equation}
p_{\rm esc} = p_{\rm knee}\left(\frac{R_{\rm sh}}{R_{\rm Sedov}} \right)^{-\alpha}~~,
\label{eq:pesc2}
\end{equation}
where $R_{\rm Sedov}$ is the SNR radius at $t=t_{\rm Sedov}$.
We assume $\alpha=6.5$ in order to $p_{\rm esc}=1~{\rm GeV}/c$ at the end of the Sedov phase \citep{ohira10}. 
From equations.~(\ref{eq:pesc1}) and (\ref{eq:pesc2}), we obtain
\begin{equation}
D_0 = \frac{\kappa R_{\rm sh}u_{\rm sh}}{p_{\rm knee}}  \left(\frac{R_{\rm sh}}{R_{\rm Sedov}} \right)^{\alpha}~~.
\label{eq:d0}
\end{equation}
Moreover, we assume that the number of CRs in the momentum range $(m_{\rm p}c,m_{\rm p}c+{\rm d}p)$ 
in the SNR is $K(R_{\rm sh}){\rm d}p\propto R_{\rm sh}^{\beta}$ where $m_{\rm p}$ is the proton mass,
and that the CR spectrum at the shock front is $p^{-s}$.
Then, in most of the cases, $N_{\rm esc}(p)$ has a single power-law form as
\begin{equation}
N_{\rm esc}(p) \propto p^{-(s+\beta/\alpha)}~~.
\label{eq:nesc1}
\end{equation}
The value of $\beta$ depends on the injection model and ranges from  $-3/4$ to $3/13$ \citep{ohira10}.
From equation~(\ref{eq:pesc2}), we derive
\begin{equation}
R_{\rm esc}(p) = (1+\kappa)R_{\rm Sedov}\left(\frac{p}{p_{\rm knee}} \right)^{-1/\alpha}~~.
\label{eq:resc}
\end{equation}
Using the Sedov solution and equation~(\ref{eq:pesc2}), one finds
\begin{equation}
t_{\rm esc}(p) = t_{\rm Sedov}\left(\frac{p}{p_{\rm knee}} \right)^{-5/2\alpha}~~.
\label{eq:tesc}
\end{equation}
In this model, $R_{\rm esc}(p)$, $t_{\rm esc}(p)$ and $N_{\rm esc}(p)$ obey single power laws, respectively, so that $F(p)$ is simply given by equations~(\ref{eq:f}) or (\ref{eq:f1}).
For $p_{\rm br,ext}>p_{\rm br,2}$,
\begin{eqnarray}
F(p)\propto \left\{ \begin{array}{ll}
p^{-(1.5\delta+s+\beta/\alpha)}\Delta t(p)^{-3/2}& (p>p_{\rm br,ext}) \\
p^{-(0.5\delta+s+(\beta-1)/\alpha)} \Delta t(p)^{-1/2}& (p_{\rm br,ext}>p>p_{\rm br,2}) \\
p^{-(s+\beta/\alpha)}  & (p_{\rm br,2}>p> p_{\rm cut}) \\
0  & (p_{\rm cut}> p)
\end{array} \right. ~~.
\end{eqnarray}
For $p_{\rm br,2}>p_{\rm br,ext}$,
\begin{eqnarray}
F(p)\propto \left\{ \begin{array}{ll}
p^{-(1.5\delta+s+\beta/\alpha)}\Delta t(p)^{-3/2}& (p>p_{\rm br,2}) \\
p^{-(s+\beta/\alpha)}  & (p_{\rm br,2}>p> p_{\rm cut}) \\
0  & (p_{\rm cut}> p)
\end{array} \right. ~~,
\end{eqnarray}
where $\Delta t(p) = t-t_{\rm esc}(p)$.

\section[]{Model 2: SNR embedded in a molecular cloud}
\label{sec:4}

In this section, we consider the case in which the whole SNR shell interacts with the molecular clouds or cavity wall which are located at $L_1<r<L_2$.
Once the CRs encounter the molecular clouds, they all are expected to escape from the SNR because  the high-density neutral gas damps plasma waves.
In this case, the functional forms of $R_{\rm esc}(p)$ and $N_{\rm esc}(p)$ are different from those given in the previous section.
At an early epoch, the SNR does not interact with the molecular clouds and $p_{\rm esc}$ is again determined by equation~(\ref{eq:pesc0}) with $\ell_{\rm esc}=\kappa R_{\rm sh}$.
If the shock front gets closer to the molecular clouds and $L_1-R_{\rm sh}<\kappa R_{\rm sh}$, $\ell_{\rm esc}$ should be replaced with $L_1-R_{\rm sh}$.
Therefore,
\begin{equation}
\ell_{\rm esc} = \min \left[\kappa R_{\rm sh}, L_1-R_{\rm sh}\right]~~.
\label{eq:ell}
\end{equation}
From equation~(\ref{eq:pesc0}), (\ref{eq:d0}) and (\ref{eq:ell}), we derive
\begin{equation}
p_{\rm esc} = p_{\rm knee} \left(\frac{R_{\rm sh}}{R_{\rm Sedov}} \right)^{-\alpha}\min \left[1, \frac{1}{\kappa}\left(\frac{L_1}{R_{\rm sh}}-1\right)\right]~~.
\label{eq:pesc3}
\end{equation}
The effective power law index of $p_{\rm esc}$ at $R_{\rm sh}=L_1/(1+\kappa)$ is
\begin{equation}
-\left. \frac{{\rm d} \log p_{\rm esc}}{{\rm d} \log R_{\rm sh}}\right |_{R_{\rm sh} \rightarrow L_1/(1+\kappa)} 
=\alpha + 1 +\frac{1}{\kappa}~~.
\end{equation}
Since $\kappa=0.04$ is small, 
this index is much larger than $\alpha$.
This means that for $R_{\rm sh}>L_1/(1+\kappa)$, $p_{\rm esc}$  rapidly decreases.
In other words, CRs with $p<p_{\rm br,1}$ escape at the same time at $t=t_{\rm Sedov}(L_1/(1+\kappa)R_{\rm Sedov})^{5/2}$ and $p_{\rm br,1}$ is given by
\begin{equation}
p_{\rm br,1} = p_{\rm knee} \left(\frac{L_1}{R_{\rm Sedov}(1+\kappa)}\right)^{-\alpha}~~.
\label{eq:pbr12}
\end{equation}
From equation (\ref{eq:pesc3}), we find
\begin{eqnarray}
R_{\rm esc}(p)= \left\{ \begin{array}{ll}
(1+\kappa)R_{\rm Sedov}\left(p/p_{\rm knee} \right)^{-1/\alpha} & (p>p_{\rm br,1}) \\
L_1   & (p\leq p_{\rm br,1}) 
\end{array} \right. ~~.
\label{eq:resc2}
\end{eqnarray}
Using the Sedov solution and Eq.~(\ref{eq:pesc3}), we find
\begin{eqnarray}
t_{\rm esc}(p)\approx \left\{ \begin{array}{ll}
t_{\rm Sedov}\left(p/p_{\rm knee} \right)^{-5/2\alpha} & (p>p_{\rm br,1}) \\
t_{\rm Sedov}\left(p_{\rm br,1}/p_{\rm knee} \right)^{-5/2\alpha}   & (p\leq p_{\rm br,1}) 
\end{array} \right. ~~.
\label{eq:tesc2}
\end{eqnarray}
For $p<p_{\rm br,1}$, the spectrum of runaway CRs is the same as CR spectrum at the shock because the CRs escape almost at the same time.
Therefore, in this model, $N_{\rm esc}$ is approximately a broken power law,
\begin{eqnarray}
N_{\rm esc}(p)\propto \left\{ \begin{array}{ll}
p^{-\left(s+\beta/\alpha \right)} & (p>p_{\rm br,1}) \\
p_{\rm br,1}^{-\beta/\alpha}p^{-s}   & (p\leq p_{\rm br,1}) 
\end{array} \right. ~~.
\label{eq:nesc2}
\end{eqnarray}
Generally speaking, the spectrum is complicated due to four breaks. However, in our cases, almost all the CRs can reach the molecular cloud, where $F(p)$ essentially has three breaks at $p_{\rm br,ext}$, $p_{\rm br,1}$ and $p_{\rm br,2}$.
If $p_{\rm br,2}<p<p_{\rm br,ext}$, then $F(p)/N_{\rm esc}(p)$ has a spectral break due 
to the break of $R_{\rm d}(p)R_{\rm esc}(p)$ [see equation~(\ref{eq:f})].
Keeping these in mind, we derive for $p_{\rm br,ext}>p_{\rm br,1}>p_{\rm br,2}$
\begin{eqnarray}
F(p)\propto \left\{ \begin{array}{ll} 
p^{-(1.5\delta+s+\beta/\alpha)} \Delta t(p)^{-3/2}& (p_{\rm br,ext}<p) \\
p^{-\left(0.5\delta+s+\left(\beta-1\right)/\alpha\right)} \Delta t(p)^{-1/2}  & (p_{\rm br,1}<p< p_{\rm br,ext}) \\
p^{-(0.5\delta+s) }   \Delta t(p)^{-1/2} & (p_{\rm br,2}<p< p_{\rm br,1}) \\
p^{-s}  & (p< p_{\rm br,2})
\end{array} \right.,
\end{eqnarray}
for $p_{\rm br,1}>p_{\rm br,ext}>p_{\rm br,2}$, 
\begin{eqnarray}
F(p)\propto \left\{ \begin{array}{ll}
p^{-(1.5\delta+s+\beta/\alpha)}  \Delta t(p)^{-3/2} & (p_{\rm br,1}<p) \\
p^{-(1.5\delta+s) }  \Delta t(p)^{-3/2} & (p_{\rm br,ext}<p< p_{\rm br,1}) \\
p^{-(0.5\delta+s) }  \Delta t(p)^{-1/2} & (p_{\rm br,2}<p< p_{\rm br,ext}) \\
p^{-s}  & (p< p_{\rm br,2})
\end{array} \right. ,
\end{eqnarray}
for $p_{\rm br,1}>p_{\rm br,2}>p_{\rm br,ext}$, 
\begin{eqnarray}
F(p)\propto \left\{ \begin{array}{ll}
p^{-(1.5\delta+s+\beta/\alpha)}  \Delta t(p)^{-3/2} & (p_{\rm br,1}<p) \\
p^{-(1.5\delta+s)}  \Delta t(p)^{-3/2} & (p_{\rm br,2}<p<p_{\rm br,1}) \\
p^{-s}  & (p< p_{\rm br,2})
\end{array} \right. ,
\end{eqnarray}
for $p_{\rm br,ext}>p_{\rm br,2}>p_{\rm br,1}$, 
\begin{eqnarray}
F(p)\propto \left\{ \begin{array}{ll}
p^{-(1.5\delta+s+\beta/\alpha)}  \Delta t(p)^{-3/2} & (p_{\rm br,ext}<p) \\
p^{-\left(0.5\delta+s+\left(\beta-1\right)/\alpha\right)} \Delta t(p)^{-1/2} & (p_{\rm br,2}<p< p_{\rm br,ext}) \\
p^{-(s+\beta/\alpha) }  & (p_{\rm br,1}<p< p_{\rm br,2}) \\
p^{-s}  & (p< p_{\rm br,1})
\end{array} \right.,
\end{eqnarray}
for $p_{\rm br,2}>p_{\rm br,ext}>p_{\rm br,1}$ and for $p_{\rm br,2}>p_{\rm br,1}>p_{\rm br,ext}$, 
\begin{eqnarray}
F(p)\propto \left\{ \begin{array}{ll}
p^{-(1.5\delta+s+\beta/\alpha)}  \Delta t(p)^{-3/2} & (p_{\rm br,2}<p)\\
p^{-(s+\beta/\alpha) }  & (p_{\rm br,1}<p< p_{\rm br,2}) \\
p^{-s}  & (p< p_{\rm br,1})
\end{array} \right..
\end{eqnarray}
Note that when $p_{\rm br,1}$ is small enough, the model 2 is the same as model 1 except for $p_{\rm cut}$ (compare equations~(23) with (35), or (24) with (36)).
\section{Gamma-ray spectrum from a SNR}
\label{sec:5}
In this section, specifying model parameters, we calculate the gamma-ray spectrum from a SNR interacting with molecular clouds and compare with observed spectra.

We consider model~2, because observed spectra have no low-energy cutoff.
From equations (\ref{eq:rd}), (\ref{eq:fp}), (\ref{eq:dism}), (\ref{eq:pbr12}), (\ref{eq:resc2}),
(\ref{eq:tesc2}), and (\ref{eq:nesc2}), one can calculate the CR momentum spectrum.
For simplicity, we assume that the injection model is the thermal leakage model, that is, $\beta = 3(3-s)/2$ \citep{ohira10}, and $R_{\rm Sedov}=2.1~{\rm pc}$ and $t_{\rm Sedov}=210~{\rm yr}$ for all SNRs.
Using the code provided by \citet{Kam+06} (see also \citet{karlsson08}), we calculate the spectrum of $\pi^0$-decay gamma rays.
The normalization of the gamma-ray flux is adjusted to fit the data.

\begin{figure}
\begin{center}
\includegraphics[width=80mm]{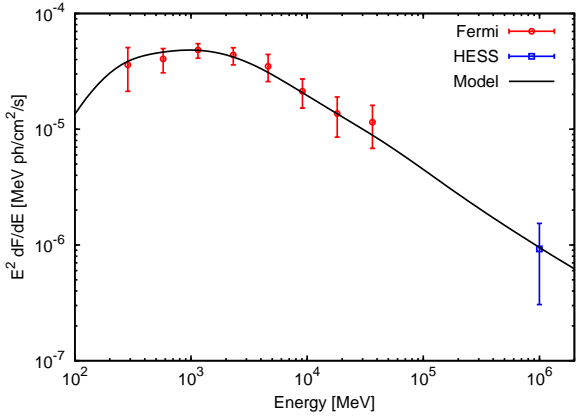}
\end{center}
\caption{Comparison of the model results (solid line) with {\it Fermi} (red) \citep{abdo09b} and HESS (blue) \citep{fiasson09} observations for SNR W51C.}
\label{fig1}
\end{figure}

\begin{figure}
\begin{center}
\includegraphics[width=80mm]{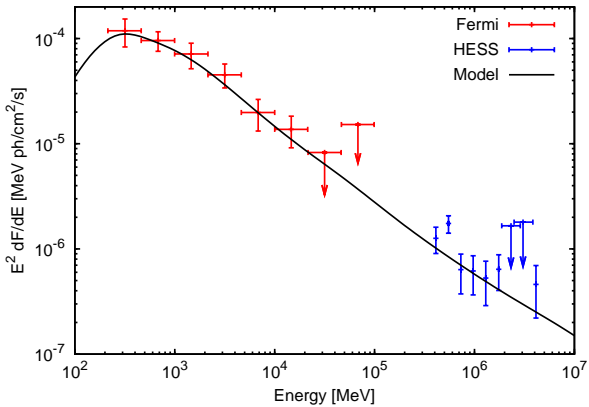}
\end{center}
\caption{Comparison of the model results (solid line) with {\it Fermi} (red) \citep{abdo10b} and HESS (blue) \citep{aharonian08} observations for the source N of SNR W28.}
\label{fig2}
\end{figure}

\begin{figure}
\begin{center}
\includegraphics[width=80mm]{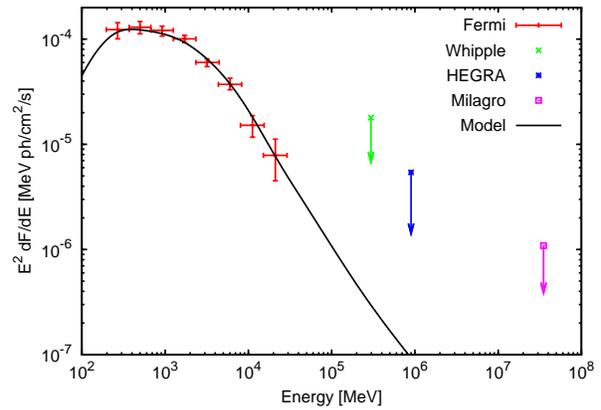}
\end{center}
\caption{Comparison of the model results (solid line) with {\it Fermi} (red) \citep{abdo10c},  {\it Whipple} (green) \citep{buckley98}, {\it HEGRA} (blue) \citep{aharonian02}, {\it Milagro} (purple) \citep{abdo09a} observations for SNR W44. 
The data of {\it Whipple}, {\it HEGRA} and {\it Milagro} are upper limits.}
\label{fig3}
\end{figure}

\begin{figure}
\begin{center}
\includegraphics[width=80mm]{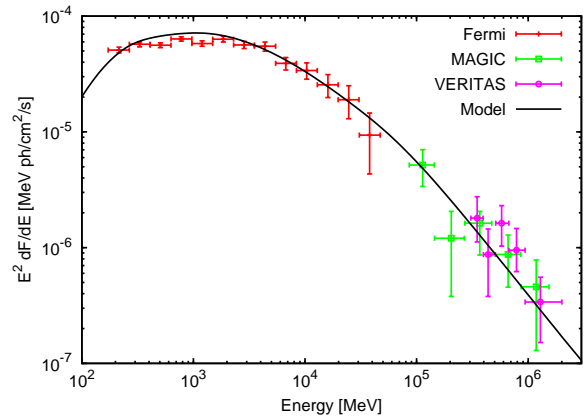}
\end{center}
\caption{Comparison of the model results (solid line) with {\it Fermi} (red) \citep{abdo10a}, {\it MAGIC} (green squares) \citep{albert07} and {\it VERITAS} (purple circles) \citep{acciari09} observations for SNR IC~443.}
\label{fig4}
\end{figure}

Figures~\ref{fig1}, \ref{fig2}, \ref{fig3} and \ref{fig4} are the results for W51C, W28, W44 and IC~443, respectively.
Table~1 is the list of adopted parameters and calculated break momentums.
We find that for all SNRs, the observed gamma-ray breaks can be attributed to the momentum break at $p_{\rm br,1}$.
For W51C and IC~443, the model predicts  second breaks seen around 30~GeV and 15~GeV, respectively (see figures~\ref{fig1} and \ref{fig4}).
For W51C, the break comes from the momentum break at $p_{\rm br,2}$ that is the effect of a finite-size emission zone, and for IC~443, the break comes from the momentum break at $p_{\rm br,ext}$ that is the effect of a finite-size source. 
Note that $p_{\rm br,2}$ was also predicted by \citet{aharonian96}, while other two breaks , $p_{\rm br,1}$ and $p_{\rm br,ext}$, are first introduced in this paper.
For W28 and W44 (figures~\ref{fig2} and \ref{fig3}), $p_{\rm br,2}$ and $p_{\rm br,ext}$ are much smaller than the threshold of the $\pi^0$ production, so that the spectrum is the same as that of the point source solution (equation~\ref{eq:fpoint}).
Our model seems consistent with observations for all the SNRs. The observed data are fitted with the same parameters concerning the escape and the acceleration, $\alpha$, $\beta$, $\kappa$, and $s$.
Hence, all the SNRs could have the same source spectrum above $p_{\rm br,1}(\sim 1-10~{\rm GeV})$, required for explaining Galactic CRs.
The observed diversity of the gamma-ray spectra comes from five parameters related to CR propagation around SNRs ($\chi$ and $\delta$) and/or environments ($L_1$, $L_2$, and $t_{\rm age}$).

\begin{table*}
 \centering
 \begin{minipage}{140mm}
  \caption{Model parameters and characteristic momentums}
  \begin{tabular}{lcccccccccccc}
  \hline
   SNR     &  $\alpha$ & $\beta$ & $\kappa$ & $s$ & $\delta$ & $\chi$ & $L_1$ & $L_2$ & $t_{\rm age}$ & $p_{\rm br,1}$ & $p_{\rm br,2}$ & $p_{\rm br,ext}$\\
 &&&&&&&(pc)&(pc)&(kyr)&(GeV/$c$)&(GeV/$c$)&(GeV/$c$)  \\
 \hline
 W51C &6.5 & 1.2 & 0.04 & 2.2 & 0.22 & 0.1  & 14.7 & 23.7 & 31.5 & 13.1 & 124         & 21.6\\
 W28   &6.5 & 1.2 & 0.04 & 2.2 & 0.19 & 0.9   & 18.5 & 26.9 & 63.0 & 2.96 & $<1.43$ & $<1.43$ \\
 W44   &6.5 & 1.2 & 0.04 & 2.2 & 0.40 & 1.0   & 12.4 & 16.2 & 23.1 & 39.8 & $<1.43$ & $<1.43$\\
 IC 443&6.5 & 1.2 & 0.04 & 2.2 & 0.62 & 0.01 & 11.5 & 14.7 & 23.1 & 62.9 & 2.93       & 151\\
 \hline
\end{tabular}
\end{minipage}
\end{table*}


\section{Discussion}
For the four SNRs detected by \textit{Fermi}, we have shown that they can be explained by an accelerated-CR spectrum with $s \approx 2.2$ and the corresponding runaway-CR spectrum with $s + \beta / \alpha \approx 2.38$. 
It is found that $s=2.1-2.4$ also gives acceptable fits.
Those spectral indices are consistent with those of the source spectrum of Galactic CRs, which is implied from GALPROP \citep{strong98,strong00} and others \citep[e.g.,][]{shibata06,putze09}, within uncertainties in the CR propagation. 
The inferred values are somewhat steeper than that predicted by the classical DSA ($s=2$), and several mechanisms have been proposed to explain such steeper spectra\citep{kirk96,zirakashvili08,ohira09b,ohira09c,fujita09}. 
However, our results are based on several approximations, and we have investigated only four middle-aged SNRs, so that results on indices are not conclusive yet. 
Further theoretical and observational studies are obviously required in order to discuss underlying acceleration mechanisms and reveal the link between those middle-aged SNRs and observed Galactic CRs.

In order to fit the data, the parameters concerning the diffusion coefficient of CRs just around the SNR, $\chi$ and $\delta$, are
different from galactic mean values, $\chi\sim1$ and $\delta\sim0.5$ (see Table~1).
Possible explanation of these discrepancies is that the runaway CRs amplify the magnetic fluctuation, so that the diffusion coefficient in the medium with the high CR density can be different from the galactic average.
The higher the CR density is, the more turbulent magnetic field, and the more frequent the scattering of CRs, resulting the smaller diffusion coefficient.
On the other hand, the magnetic turbulence becomes weak if the neutral particles impedes the magnetic-field amplification.
Further studies are needed to understand physics of partially ionized interstellar matter illuminated by CRs.

Some important processes neglected so far should be mentioned.
At first, in \S~\ref{sec:2}, the diffusion coefficient of CRs in the interstellar medium, $D_{\rm ISM}(p)$, is assumed to be spatially uniform and time-independent. 
\citet{fujita10} demonstrated that CRs that have escaped from an SNR excite plasma waves and reduce the diffusion coefficient \citep{wentzel69, kulsrud69,ptuskin08}, which is responsible for non-uniform, time-dependent diffusion coefficient.
Secondly, in \S~\ref{sec:3} and \S~\ref{sec:4},
we assume the diffusion coefficient just around the shock has a form of Eq.~(\ref{eq:d0}).
Although the magnetic field amplification around the shock is studied by the linear analysis 
\citep{bell04,reville07,ohira09b} and many numerical simulations 
\citep{lucek00,giacalone07,niemiec08,reville08,requelme09,inoue09,ohira09a,VBE09}, 
the saturation of the magnetic-field amplification and the diffusion coefficient has not yet been understood in detail.
Finally, in \S~\ref{sec:3} and \S~\ref{sec:4}, we assume that the CR spectrum at the shock has a single power law, $p^{-s}$, at any time.
However, the situation may not be so simple.
The shock compression ratio decreases as the shock velocity decreases.
As a result, the spectrum becomes steeper with time \citep{fujita09}.
In addition, neutral particles have great impacts on the magnetic field and the particle acceleration \citep{ohira09b,ohira09c}. 
Moreover, nonlinear DSA should be considered when CRs affect the shock structure \citep{drury81,malkov01}, along with properties of magnetic turbulence in magnetic field amplification~\citep{VBE09}.
Further studies are needed to clarify the significance of these effects.

\section*{Acknowledgments}

We thank Y. Fujita, F. Takahara, K. Kohri and T. Kamae for useful discussion.
This work is supported in part by grant-in-aid from the Ministry of Education,
Culture, Sports, Science, and Technology (MEXT) of Japan, 
No.~21684014 (Y.~O.), No.~19047004, No.~21740184, No.~21540259 (R.~Y.).


\label{lastpage}

\begin{thebibliography}{99}

\bibitem[\protect\citeauthoryear{Abdo et al.}{2009a}]{abdo09a} Abdo, A. A., et al., 2009a, ApJ, 700, L127

\bibitem[\protect\citeauthoryear{Abdo et al.}{2009b}]{abdo09b} Abdo, A. A., et al., 2009b, ApJ, 706, L1

\bibitem[\protect\citeauthoryear{Abdo et al.}{2010a}]{abdo10a} Abdo, A. A., et al., 2010a, ApJ, 712, 459

\bibitem[\protect\citeauthoryear{Abdo et al.}{2010b}]{abdo10b} Abdo, A. A., et al., 2010b, ApJ, 718, 348

\bibitem[\protect\citeauthoryear{Abdo et al.}{2010c}]{abdo10c} Abdo, A. A., et al., 2010c, Science, 327, 1103

\bibitem[\protect\citeauthoryear{Acciari et al.}{2009}]{acciari09} Acciari, V. A., et al., 2009, ApJ, 698, L133

\bibitem[\protect\citeauthoryear{Aharonian \& Atoyan}{1996}]{aharonian96} Aharonian, F. A., \& Atoyan, A., 1996, A\&A, 309, 917

\bibitem[\protect\citeauthoryear{Aharonian et al.}{2002}]{aharonian02} Aharonian, F. A., et al., 2002, A\&A, 395, 803

\bibitem[\protect\citeauthoryear{Aharonian et al.}{2008}]{aharonian08} Aharonian, F. A., et al., 2008, A\&A, 481, 401

\bibitem[\protect\citeauthoryear{Albert et al.}{2007}]{albert07} Albert, J., et al, 2007 ApJ, 664, L87

\bibitem[\protect\citeauthoryear{Atoyan et al.}{1995}]{atoyan95} Atoyan, A. M., Aharonian, F. A. \& V{\"o}lk, H. J., 1995, PRD, 52, 3265

\bibitem[\protect\citeauthoryear{Axford et al.}{1977}]{axford77}Axford, W. I., Leer, E., \& Skadron, G., 1977, Proc. 15th Int. Cosmic Ray Conf., Plovdiv, 11, 132

\bibitem[\protect\citeauthoryear{Baade \& Zwicky}{1934}]{baade34}Baade, W., \& Zwicky, F., 1934, Proc. Natl. Acad. Sci., 20, 259

\bibitem[\protect\citeauthoryear{Bell}{1978}]{bell78}Bell, A. R., 1978, MNRAS, 182, 147

\bibitem[\protect\citeauthoryear{Bell}{2004}]{bell04}Bell, A. R., 2004, MNRAS, 353, 550 

\bibitem[\protect\citeauthoryear{Berezinskii et al.}{1990}]{berezinskii90} Berezinskii, V. S., Bulanov, S. V., Dogiel, V. A., Ginzburg, V. L., Ptuskin, V. S., 1990, Astrophysics of Cosmic Rays. North Holland, Amsterdam

\bibitem[\protect\citeauthoryear{Blandford \& Ostriker}{1978}]{blandford78}Blandford, R. D., \& Ostriker, J. P., 1978, ApJ, 221, L29

\bibitem[\protect\citeauthoryear{Buckley et al.}{1998}]{buckley98}Buckley, J. H., et al., 1998, A\&A, 329, 639

\bibitem[\protect\citeauthoryear{Drury \& V\"{o}lk}{1981}]{drury81}Drury, L.O'C., \& V\"{o}lk, H. J. 1981, ApJ, 248, 344

\bibitem[\protect\citeauthoryear{Fiasson et al.}{2009}]{fiasson09}Fiasson, A., Marandon, V., Chaves, R. J.G., \& Tibolla, O. for the H.E.S.S. collaboration, 2009, in Proc. 31st Int. Cosmic-Ray Conf., in press

\bibitem[\protect\citeauthoryear{Fujita et al.}{2009}]{fujita09}Fujita, Y., Ohira, Y., Tanaka, S. J., \& Takahara, F., 2009, ApJ, 707, L179

\bibitem[\protect\citeauthoryear{Fujita et al.}{2010}]{fujita10}Fujita, Y., Ohira, Y., \& Takahara, F., 2010, ApJ, 712, L153

\bibitem[\protect\citeauthoryear{Gabici et al.}{2009}]{gabici09}
Gabici, S., Aharonian, F. A., \& Casanova, S. 2009, MNRAS, 369, 1629

\bibitem[\protect\citeauthoryear{Giacalone \& Jokipii}{2007}]{giacalone07} Giacalone, J., \& Jokipii, J.R., 2007, ApJ, 663, L41

\bibitem[\protect\citeauthoryear{Giuliani et al.}{2010}]{giuliani10}Giuliani, et al., 2010, preprint (arXiv:1005.0784)

\bibitem[\protect\citeauthoryear{Inoue et al.}{2009}]{inoue09} Inoue, T., Yamazaki, R. \& Inutsuka, S., 2009, ApJ, 695, 825

\bibitem[\protect\citeauthoryear{Kamae et al.}{2006}]{Kam+06}Kamae, T., et al., 2006, ApJ,  647, 692

\bibitem[\protect\citeauthoryear{Karlsson \& Kamae}{2008}]{karlsson08}Karlsson, N., \& Kamae, T., 2008, ApJ, 674, 278 

\bibitem[\protect\citeauthoryear{Kirk et al.}{1996}]{kirk96}Kirk, J. G., Duffy, P., and Gallant, Y.A., 1996, A\&A, 314, 1010

\bibitem[\protect\citeauthoryear{Kulsrud \& Pearce}{1969}]{kulsrud69} Kulsrud, R. M., \& Pearce, W. P.,1969, ApJ, 156, 445

\bibitem[\protect\citeauthoryear{Krymsky}{1977}]{krymsky77} Krymsky, G. F., 1977, Dokl. Akad. Nauk SSSR, 234, 1306

\bibitem[\protect\citeauthoryear{Lucek \& Bell}{2000}]{lucek00} Lucek, S. G. \& Bell, A. R. 2000, MNRAS, 314, 65

\bibitem[\protect\citeauthoryear{Malkov \& Drury}{2001}]{malkov01}Malkov, M. A., \& Drury, L.O'C.\ 2001, Rep.\ Prog.\ Phys., 64, 429

\bibitem[\protect\citeauthoryear{Malkov et al.}{2010}]{malkov10}Malkov, M. A., Diamond, P. H., \& Sagdeev, R. Z., preprint (arXiv:1004.4714)

\bibitem[\protect\citeauthoryear{Niemiec et al.}{2008}]{niemiec08} Niemiec, J., Polh, M., \& Nishikawa, K., 2008, ApJ, 684, 1174

\bibitem[\protect\citeauthoryear{Ohira et al.}{2009a}]{ohira09a} Ohira, Y., Reville, B., Kirk, J. G., \& Takahara, F., 2009a, ApJ, 698, 445

\bibitem[\protect\citeauthoryear{Ohira et al.}{2009b}]{ohira09b}Ohira, Y., Terasawa, T., \& Takahara, F., 2009b, ApJ, 703, L59

\bibitem[\protect\citeauthoryear{Ohira et al.}{2009c}]{ohira09c}Ohira, Y., \& Takahara, F., 2009c, preprint (arXiv:0912.2859)

\bibitem[\protect\citeauthoryear{Ohira et al.}{2010}]{ohira10}Ohira, Y., Murase, K., \& Yamazaki, R., 2010, A\&A, 513, A17

\bibitem[\protect\citeauthoryear{Ptuskin et al.}{2008}]{ptuskin08}Ptuskin, V. S., Zirakashvili, V. N., \& Plesser, A. A., 2008, Adv. Space Res., 42, 486 755

\bibitem[\protect\citeauthoryear{Ptuskin \& Zirakashvili}{2005}]{ptuskin05}Ptuskin, V. S., \& Zirakashvili, V. N., 2005, A\&A, 429, 755

\bibitem[\protect\citeauthoryear{Putze et al.}{2009}]{putze09} Putze, A., Derome, L., Maurin, D., Perotto, L., \& Taillet, R. 2009, A\&A, 497, 991

\bibitem[\protect\citeauthoryear{Riquelme \& Spitkovsky}{2009}]{requelme09}Riquelme, M. A., \& Spitkovsky, A. 2009, ApJ, 694, 626

\bibitem[\protect\citeauthoryear{Reville et al.}{2007}]{reville07} Reville, B., Kirk, J. G., Duffy, P. \& O'Sullivan, S., 2007, A\&A, 475, 435

\bibitem[\protect\citeauthoryear{Reville et al.}{2008}]{reville08} Reville, B., O'Sullivan, S., Duffy, P., \& Kirk, J. G., 2008, MNRAS, 386, 509

\bibitem[\protect\citeauthoryear{Shibata et al.}{2006}]{shibata06} Shibata, T. et al. 2006, ApJ, 642, 882

\bibitem[\protect\citeauthoryear{Strong \& Moskalenko}{1998}]{strong98} Strong, A. W., \& Moskalenko, I. V. 1998, ApJ, 509, 212

\bibitem[\protect\citeauthoryear{Strong et al.}{2000}]{strong00} Strong, A. W., Moskalenko, I. V., \& Reimer, O. 2000, ApJ, 537, 763

\bibitem[\protect\citeauthoryear{Tavani et al.}{2010}]{tavani10} Tavani, M. et al., 2010, ApJ, 710, L151

\bibitem[\protect\citeauthoryear{Vladimirov et al.}{2009}]{VBE09}Vladimirov, A.~E., Bykov, A.~M., \& Ellison, D.~C. 2009, ApJ, 703, L29

\bibitem[\protect\citeauthoryear{Weaver et al.}{1977}]{weaver77} Weaver, R., McCray, R., \& Castor, J., 1977, ApJ, 218, 377

\bibitem[\protect\citeauthoryear{Wentzel}{1969}]{wentzel69} Wentzel, D. G., 1969, ApJ, 156, 303

\bibitem[\protect\citeauthoryear{Zirakashvili \& Ptuskin}{2008}]{zirakashvili08} Zirakashvili, V. N., \& Ptuskin, V. S., 2009, in AIP Conf. Proc. 1085, High Energy Gamma-Ray Astronomy, ed. F. A. Aharonian, W. Hofmann, \& F. M. Rieger(Melville, NY: AIP), 336

\end{thebibliography}
\end{document}